\documentclass[twocolumn,showpacs,preprintnumbers,amsmath,amssymb]{revtex4}
\usepackage{psfig}
\usepackage{dcolumn}
\usepackage{bm}
\usepackage{feynmf}
\usepackage{multirow}
\begin{document}


\title{Attractive and repulsive contributions of medium fluctuations 
to nuclear superfluidity}

\author{G.Gori$^{a,b}$, F.Ramponi$^{a,b}$, F.Barranco$^{c}$, P.F. Bortignon$^{a,b}$,
R.A. Broglia$^{a,b,d}$, G. Col\`o$^{a,b}$, E.Vigezzi$^{b}$\\
$^a$ Dipartimento di Fisica, Universit\`a degli Studi di Milano,
via Celoria 16, 20133 Milano, Italy.\\
$^b$ INFN, Sezione di Milano, via Celoria 16, 20133 Milano, Italy.\\
$^c$ Departamento de Fisica Aplicada III, Escuela Superior de Ingenieros, Camino de los Descubrimientos s/n,
41092 Sevilla, Spain.\\
$^d$ The Niels Bohr Institute, University of Copenhagen, Blegdamsvej 17, 2100 Copenhagen \O, Denmark.}

\date{\today}




\begin{abstract}


Oscillations of mainly surface character ($S=0$ modes) 
give rise, in atomic nuclei, to an attractive (induced)  pairing
interaction, while spin ($S=1$) modes of mainly volume character generate a
repulsive interaction, the net effect being an attraction which accounts for a
sizeable fraction of the experimental pairing gap. Suppressing
the particle-vibration coupling mediated by the proton degrees of freedom,
i.e., mimicking neutron matter, the total surface plus spin-induced pairing interaction becomes repulsive. 

\end{abstract}

\maketitle


A central issue in a quantitative description of superconductors (metals,
doped fullerides, etc.) as well as superfluid Fermi systems ($^3$He, trapped
gases of fermionic
atoms, atomic nuclei, neutron stars, etc.) is related to the interaction 
acting between
fermions and giving rise to Cooper pairs.
The glue holding together the fermions of each Cooper pair, is the result of the bare interaction acting between fermions,
strongly renormalized by medium polarization effects: Coulomb interaction plus
plasmon and phonon exchange in metals \cite{schrieffer}, Van der Waals interaction plus spin and density
modes exchange in trapped gases of fermionic atoms \cite{pethick,Combescot}, strong force and spin and density 
modes exchange in the case of atomic nuclei and of neutron stars 
\cite{ainsworth,baldo,neut}.

While broad consensus exists concerning the mechanism of electron-electron and electron-phonon
interaction leading to superconductivity in metals, 
the situation is much less clear in the case of strongly 
interacting particles in the different scenarios found in nature. In particular,
it has been found that medium polarization effects associated with the exchange 
of vibrations lead
to a quenching of the bare pairing interaction in the $^{1}S_0$ 
channel in  the inner crust of 
neutron stars \cite{lomb}
while, at the same time, account for a sizeable fraction of the 
pairing gap in open shell nuclei \cite{PRL,Terasaki,EPJ}.
In the present paper we present, for the first time, evidence which allows
to understand these seemingly contradictory results. 

 


\begin{figure}[t]
\centerline{\psfig{file=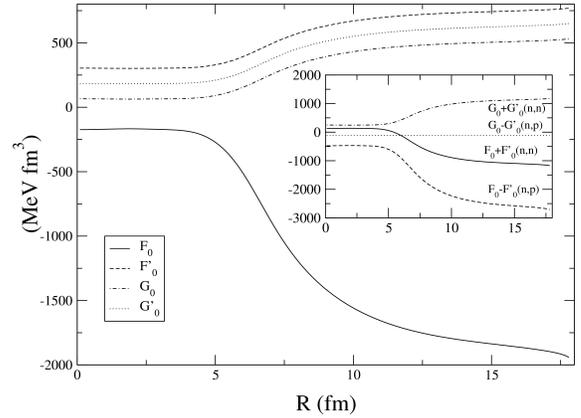,width=7.5cm}}
\caption{Generalized Landau parameters associated with the interaction SkM$^*$
defining the strength of the particle-hole interaction in the
isoscalar ($F_0$), isovector ($F'_0$), spin isoscalar ($G_0$)
and spin isovector ($G'_0$) channels. In the inset the functions
$F_0+F'_0$ (n-n interaction), 
$F_0-F'_0$ (n-p), 
$G_0+G'_0$ (n-n) and 
$G_0-G'_0$ (n-p) 
are also shown.
}
\label{fig:landp}
\end{figure}

We address the question at hand within the scenario provided
by the paradigmatic (superfluid) open shell nucleus $^{120}$Sn. 
The starting point corresponds to the calculation of the mean field
potential and associated quasiparticle properties within the framework of Hartree-Fock plus BCS theory 
\cite{HFB} using the SkM$^*$ force \cite{skm}. The polarization quanta were worked out within the 
framework of the
quasi-particle random phase approximation (QRPA). The particle-hole residual 
interaction is derived 
in a self-consistent
way from the Skyrme energy functional, with the exception of the spin-orbit and
of the Coulomb part (cf. 
\cite{QRPA} for more details). On the other hand, we neglect 
the momentum-dependent
part of the interaction in the calculation of the particle-vibration 
coupling discussed below \cite{vangiai}. The relevant part of the 
the particle-hole interaction can then be written
\begin{eqnarray}
&& v_{ph}(\vec{r},\vec{r}')= \delta(\vec{r}-\vec{r}') \times \nonumber\\
&& \times \left\{
\left[F_0+F'_0\vec{\tau}\cdot\vec{\tau}'\right] 
+\left[\left(G_0+G'_0\vec{\tau}\cdot\vec{\tau}'\right)
\vec{\sigma}\cdot\vec{\sigma}' \right]
\right\}.
\label{residual}
\end{eqnarray}
We shall only consider the $\tau_z\cdot\tau_z$ term, in keeping with the fact that we are here
interested in the neutron-neutron pairing interaction. Off-diagonal terms
are associated with charge-exchange modes. Thus, in lowest order, they do not
contribute to the neutron-neutron interaction, but are expected to be of relevance in the
discussion of the proton-neutron pairing interaction.

The functions $F_0(r)$, $F'_0(r)$, $G_0(r)$ and $G'_0(r)$
(generalized Landau-Migdal\cite{Landau,Migdal} parameters) controlling the isoscalar and isovector (spin-independent
and spin-dependent) channels are displayed in Fig. \ref{fig:landp}.

\begin{figure}[t]
\centerline{\psfig{file=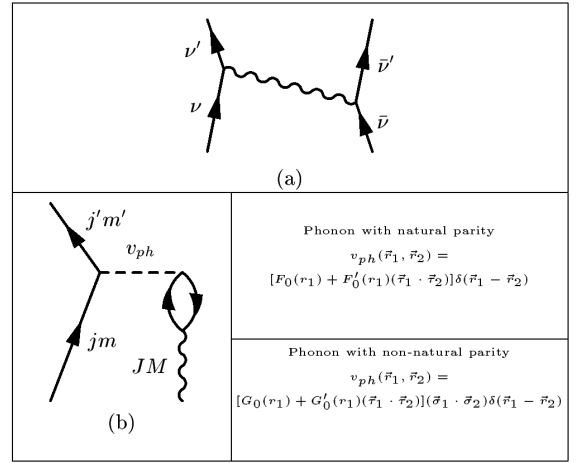,width=7.5cm}}
\caption{(a) diagram depicting the pairing interaction induced
by the exchange of phonons; (b) particle-vibration coupling vertex making it 
explicit the dominant part of the particle-hole interaction giving rise, through the sum of bubble
 diagrams, to the corresponding QRPA modes (wavy line).}
\label{fig:fgraph}
\end{figure}

Strictly speaking, in the case of atomic nuclei spin is not a good quantum number with which to identify the
polarization quanta, because of the strong spin-orbit term present in these systems. 
We have thus adopted the criterion of distinguishing between natural
($\pi = (-1)^J)$ and non-natural ($\pi = - (-1)^J$) parity modes, where $J$
indicates the total angular momentum of the quanta. 
Vibrations of multipolarity and parity $J^{\pi}=1^+$, $2^+$,$2^-$, 
$3^+$,$3^-$, $4^+$,$4^-$,$5^+$ and $5^-$ were worked out.
Those having energy $\le$30 MeV 
were used in the calculation of the induced interaction (cf. 
Fig. \ref{fig:fgraph}(a)). 
From the QRPA calculation one gets \cite{Bertsch}, together with the energy
of the excited states, their transition densities which
will be used as form factor for the particle-vibration coupling vertex
(cf. Fig. \ref{fig:fgraph}(b)):
\begin{equation}
\begin{array}{c}
\delta \rho_{J^{\pi}}^{i}(r)=\frac{1}{\sqrt{2J+1}}\sum_{\nu_1,\nu_2}
(X_{\nu_1,\nu_2}(i,J^{\pi})+Y_{\nu_1,\nu_2}(i,J^{\pi})) \\
\\
       \times  (u_{\nu_1}v_{\nu_2}+u_{\nu_2}v_{\nu_1}) \times  
       \langle \nu_1||i^JY_{J}||\nu_2 \rangle\varphi_{\nu_1}(r)\varphi_{\nu_2}(r).\\
\end{array}
\label{eqn:td-density}
\end{equation}

\begin{equation}
\begin{array}{c}
\delta\rho_{J^{\pi}L}^{i}(r)=\frac{1}{\sqrt{2J+1}}
\sum_{\nu_1,\nu_2}(X_{\nu_1,\nu_2}(i,J^{\pi})-Y_{\nu_1,\nu_2}(i,J^{\pi})) \\
\\
\times (u_{\nu_1}v_{\nu_2}+u_{\nu_2}v_{\nu_1}) \times  
\langle \nu_1||i^L[Y_L\times\sigma]_J|| \nu_2\rangle\varphi_{\nu_1}(r)
\varphi_{\nu_2}(r). \\
\end{array}
\label{eqn:td-spin}
\end{equation}

The function (\ref{eqn:td-density})
is associated with  the response to external
fields which induce a density oscillation; it vanishes for phonons of unnatural
parity. 
The function (\ref{eqn:td-spin})
is instead associated with the response
to magnetic external fields and hence the coupling to excited
states mediated by the part of the residual interaction which
depends on the spin, and it applies to phonons of both unnatural
(when $J \neq L$) and natural (when $J=L$) parity.
 The index 
$i$ labels the different vibrational modes of a given spin and parity, while $X$ and $Y$ are the forwardsgoing and backwardsgoing QRPA amplitudes
of the corresponding modes, and the index $\nu$ denotes the quantum numbers 
$n,l,j$ of the single particle states.

By using Eq. (\ref{residual}),
one can find the expression of the vertices produced by the
spin independent part of the residual interaction:
\begin{eqnarray}
&&f^{\nu' m'}_{\nu m ; J^{\pi}Mi}= \nonumber \\
&& \langle \nu'm' |
\left[F_0(r)+F'_0(r)\vec{\tau}\cdot\vec{\tau}'\right] 
\delta(\vec{r}-\vec{r}')|\nu m; J^{\pi}Mi\rangle.
\end{eqnarray}
It can be rewritten by using the multipole  expansion for the $\delta$ function as 
\begin{eqnarray}
&& f_{\nu m ; J^{\pi}Mi}^{\nu' m' } = i^{l-l'}
\langle j'm'| (i)^JY_{JM}|jm\rangle  \times \nonumber \\
&& \int dr \varphi_{\nu'}[(F_0+F_0')\delta\rho_{J^{\pi}n}^{i}
+(F_0-F_0')\delta\rho_{J^{\pi}p}^{i}] \varphi_{\nu},
\label{eqn:3}
\end{eqnarray}
$\delta\rho^i_{J^{\pi}n}$ and $\delta\rho^i_{J^{\pi}p}$ being the neutron and proton
contributions to the transition densities defined in Eq. (\ref{eqn:td-density}).
In a similar way, the vertices produced by the spin dependent part of the residual interaction are
\begin{eqnarray}
&& g_{\nu m ; J^{\pi}Mi}^{\nu' m' }=  \nonumber \\
&& \langle \nu'm'|\left[G_0(r)+G'_0(r)\vec{\tau}\cdot\vec{\tau}'\right]
\vec{\sigma}\cdot\vec{\sigma}' \delta(\vec{r}-\vec{r}')|\nu m; J^{\pi}Mi\rangle
\end{eqnarray}
and, as before, they can be expanded in the form
\begin{eqnarray}
&& g_{\nu m J^{\pi}Mi}^{\nu' m'} = \sum_{L=J-1}^{J+1}
i^{l-l'}
\langle j'm'
| (i)^L[Y_L\times\sigma]_{JM}|j m\rangle \times \nonumber \\
&& \int d r\varphi_{\nu'} 
[(G_0+G_0')\delta\rho_{J^{\pi}Ln}^{i}+(G_0-G_0')\delta\rho_{J^{\pi}Lp}^{i}]
\varphi_{\nu}.
\label{eqn:4}
\end{eqnarray}
where $\delta\rho^i_{J^{\pi}Ln}$ and $\delta\rho^i_{J^{\pi}Lp}$ are respectively 
the neutron and proton
contributions to the transition densities defined in Eq. (\ref{eqn:td-spin}).

\begin{figure}[t]
\centerline{\psfig{file=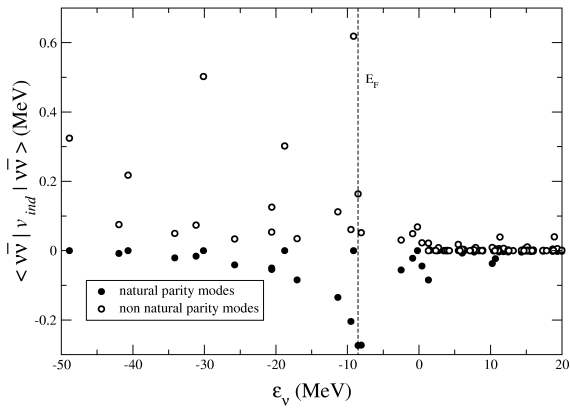,width=7.5cm}}
\caption{
Diagonal induced pairing matrix elements resulting from the exchange
of phonons with natural parity (filled circles) and those resulting from
the exchange of phonons with non-natural parity vibrations (empty circles),
displayed as a function of the energy of the single-particle
state $\epsilon_{\nu}$. 
} \label{fig:diagonal}
\end{figure}

\begin{figure}[!b]
\centerline{\psfig{file=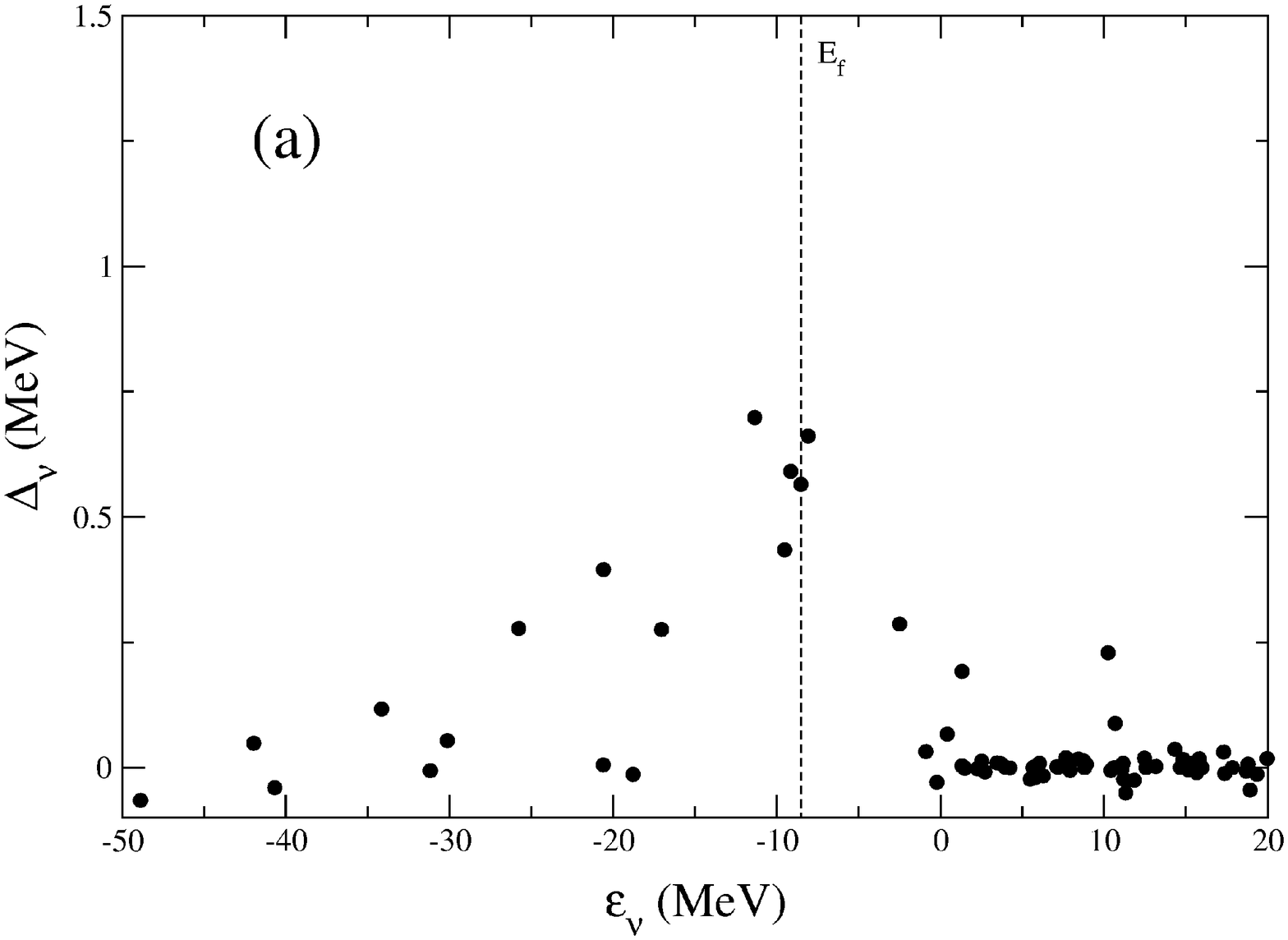,width=7.5cm}}
\vspace {0.3cm}
\centerline{\psfig{file=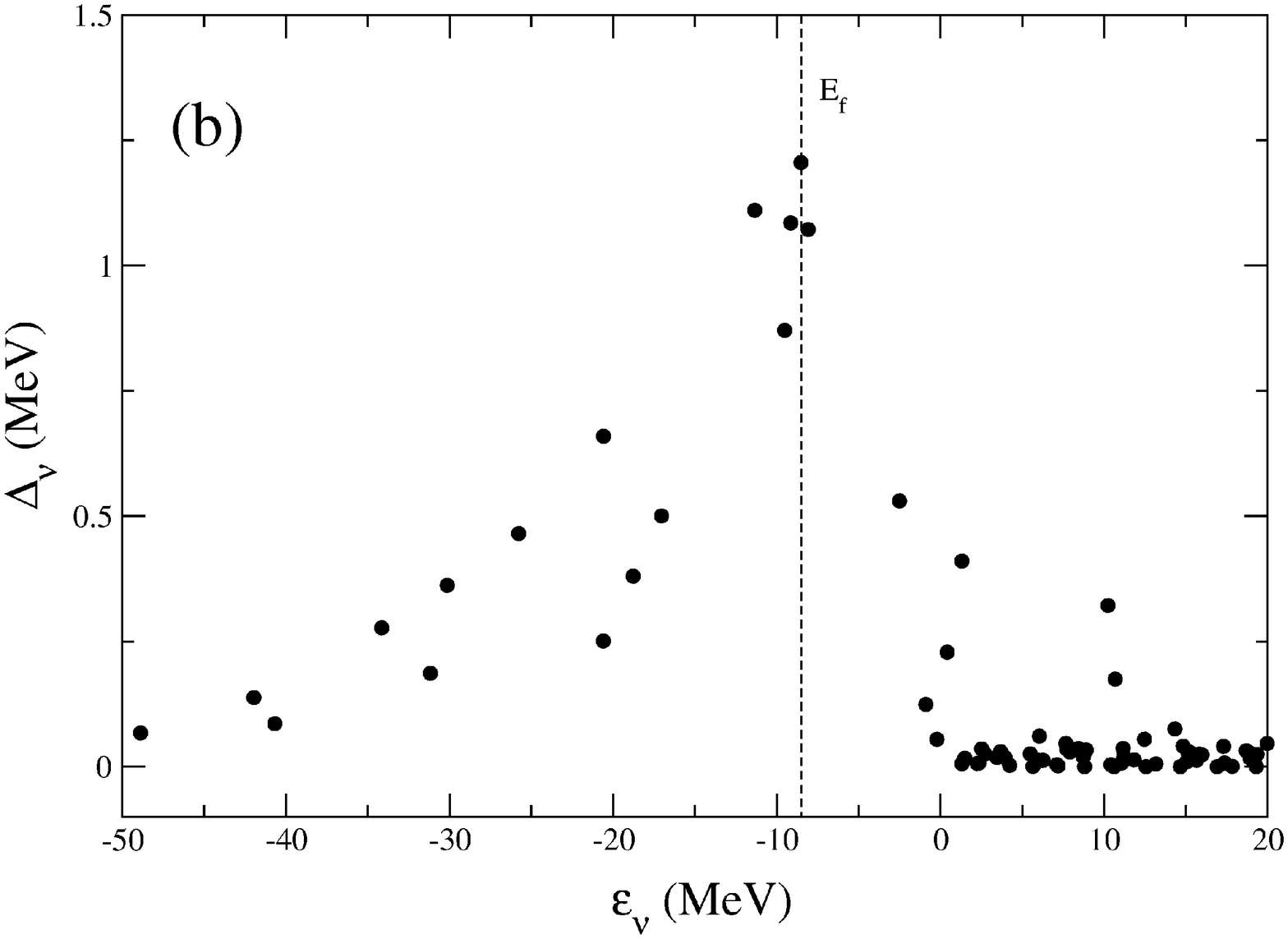,width=7.5cm}
}
\caption{
(a) 
The state dependent pairing gap as a function of the single-particle energies
obtained by solving the BCS equations associated with the total  
induced interaction matrix elements; 
(b) same as (a) but for the 
matrix elements 
produced including only the spin independent part of the particle-hole
interaction.
}
\label{fig:gap}
\end{figure}

These particle-vibration coupling matrix elements, together with the 
energies of the QRPA modes and the 
HF single-particle
energies, are the basic ingredients needed to calculate the pairing
induced interaction $v_{ind}$ (cf. Fig. \ref{fig:fgraph}(a)) within the 
framework presented in ref. \cite{PRL}.
The matrix elements between  two pairs of neutrons in time-reversal states
are given by 
\begin{eqnarray}
&& <\nu' m' \nu' \bar{m'}| v_{ind}|\nu m \nu\bar{m}> = \nonumber \\
&& \sum_{J^{\pi}Mi} 
\frac {2 (f+g)^{\nu'  m'}_{\nu  m ; J^{\pi}Mi} 
(f-g)^{\nu'  m'}_{\nu  m ; J^{\pi}Mi}}{E_0 - E_{int}} 
\label{vind}
\end{eqnarray}
where the sum is over the phonons of multipolarity ${J^{\pi},M}$
obtained in the QRPA calculation. 
In the denominator, $E_{int} = (|\epsilon_{\nu}- \epsilon_F| 
                              + |\epsilon_{\nu'}-\epsilon_F| + h\omega_i)$ 
denotes the energy of the intermediate state  
given by two particles (whose energy is calculated respect to the
Fermi energy $\epsilon_F$) and one vibration, while $E_0$ is
the energy of the correlated  two-particle state, which must be 
obtained self-consistently: the denominator is therefore always negative.
The important conclusion is that
the sign of the matrix element (\ref{vind}) then depends on the relative magnitudes of 
the attractive contribution $f^2$, arising  from the spin-independent term  
and of the repulsive contribution $-g^2$ arising from  
the spin-dependent term.

The matrix elements  (\ref{vind}) are then used to obtain the $^1S_0$ neutron pairing gap, 
solving the
BCS gap equation \cite{note}:
\begin{equation}
\Delta_{\nu} = - \frac{1}{2j+1}\sum_{\nu'}  \frac{\Delta_{\nu'}}{2 E_{\nu'}} 
v_{\nu,\nu'}
\label{gapeq}
\end{equation}
where 
\begin{eqnarray}
&& v_{\nu,\nu'} =  \sum_{mm'}  <\nu m \nu \bar{m}| v_{ind}|\nu' m' \nu'\bar{m'}> =  
\nonumber \\
&& {\sqrt{(2j+1)(2j'+1)}}  <jj;0| v_{ind} | j'j';0>   
\label{vmat}
\end{eqnarray}


The two particle wavefunction $|j'j';0>$, coupled to zero angular momentum, 
contains an admixture of  singlet and triplet   components, with about equal 
weight.  It is well known (see e.g. \cite{berk,Nakajima,peth})
that in infinite matter, when only a species is present, like in neutron
matter,
the matrix elements associated with the singlet component have a simple character, depending on the spin $S$ of the 
exchanged fluctuation: for density modes, characterized by  $S$ =0, the
matrix elements are negative, while for spin modes ($S=1$) they are positive.
The suppression of the pairing gap produced by the bare neutron-neutron force
through medium polarization effects 
in neutron matter is associated with the dominance of the repulsive 
contribution from the spin modes 
over the attractive contribution from the density modes.


In the case of $^{120}$Sn, we observe that 
the diagonal matrix elements $<jj;0| v_{ind} | jj;0> $, shown in
in Fig. \ref{fig:diagonal}, are systematically attractive or repulsive, 
depending  on whether the exchanged fluctuations are respectively of natural 
or unnatural parity, in direct correspondence with the case of infinite matter.
On one hand, this is because
for non natural parity  modes only the spin dependent vertices  (\ref{eqn:4}),
which have a $S=1$ character, contribute, while
for natural parity modes, the spin independent matrix elements ({\ref{eqn:3}}) 
are the dominant ones,
and one can show that they are the only ones contributing to 
the diagonal matrix elements.    
On the other hand, the contributions of the matrix elements 
from triplet component of the two-particle $|jj0>$ 
wavefunction are small and of variable sign \cite{bertsch}.

However, in constrast with neutron matter, the resulting total matrix elements 
are  predominantly attractive (in any case around the Fermi energy) \cite{note1}.

The resulting state dependent pairing gap obtained
by solving
the BCS gap and number equations making use of the (total) induced 
pairing matrix
elements is depicted in Fig. \ref{fig:gap}(a). For states
close to the Fermi energy the gap accounts for a consistent fraction
of the experimental value (1.4 MeV). 
If one solves the BCS 
equations considering
only the exchange of density modes (i.e. neglecting the contributions
from Eq. (\ref{eqn:4})), one obtains values 
which are, in average, larger (cf. Fig. \ref{fig:gap}(b)).
In fact, the exchange of $S=1$ modes quenches the pairing gap arising from the
exchange of only $S=0$ modes by roughly 30\% \cite{addgap}.

\begin{figure}[t]
\centerline{\psfig{file=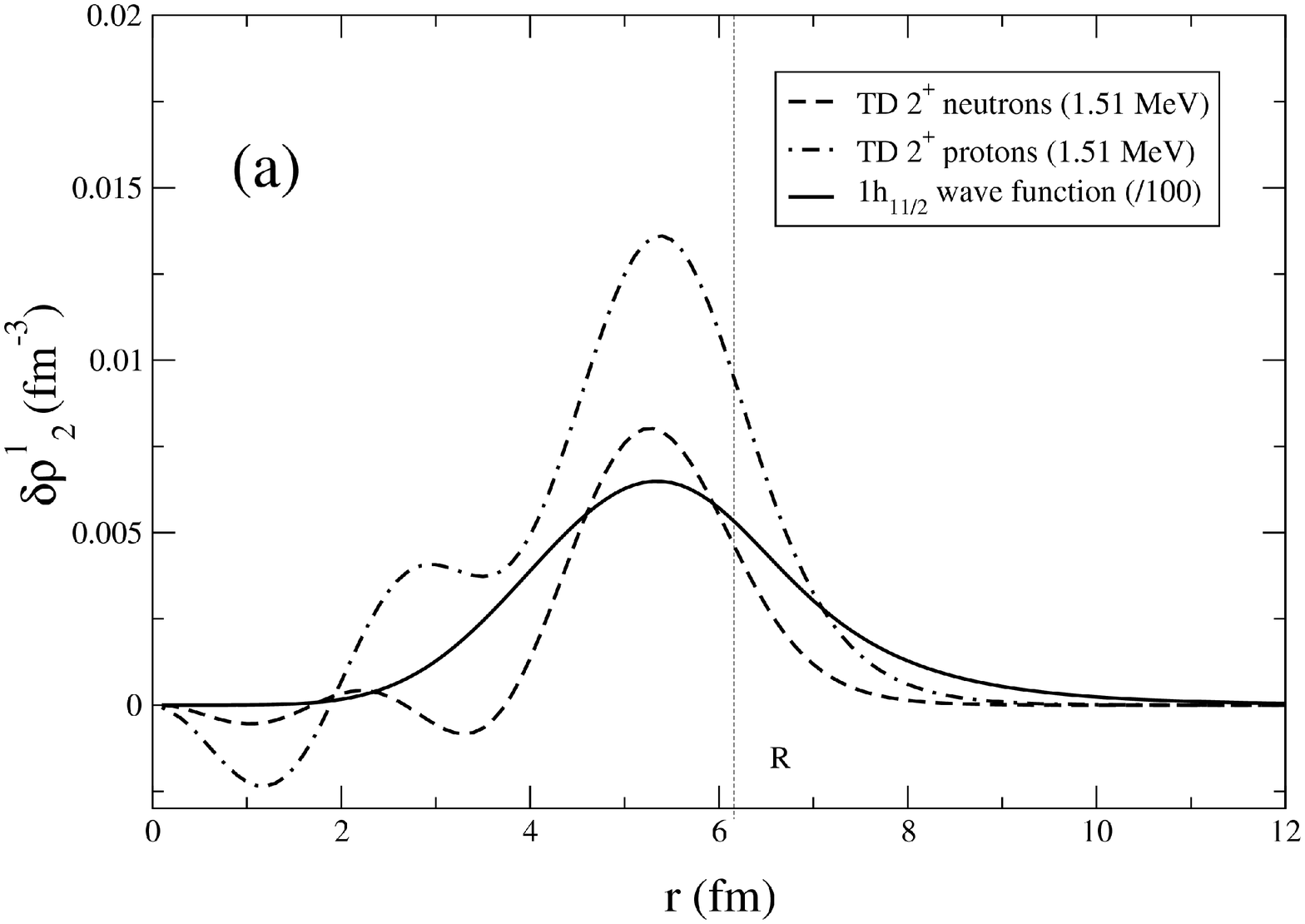,width=7.5cm}}
\centerline{\psfig{file=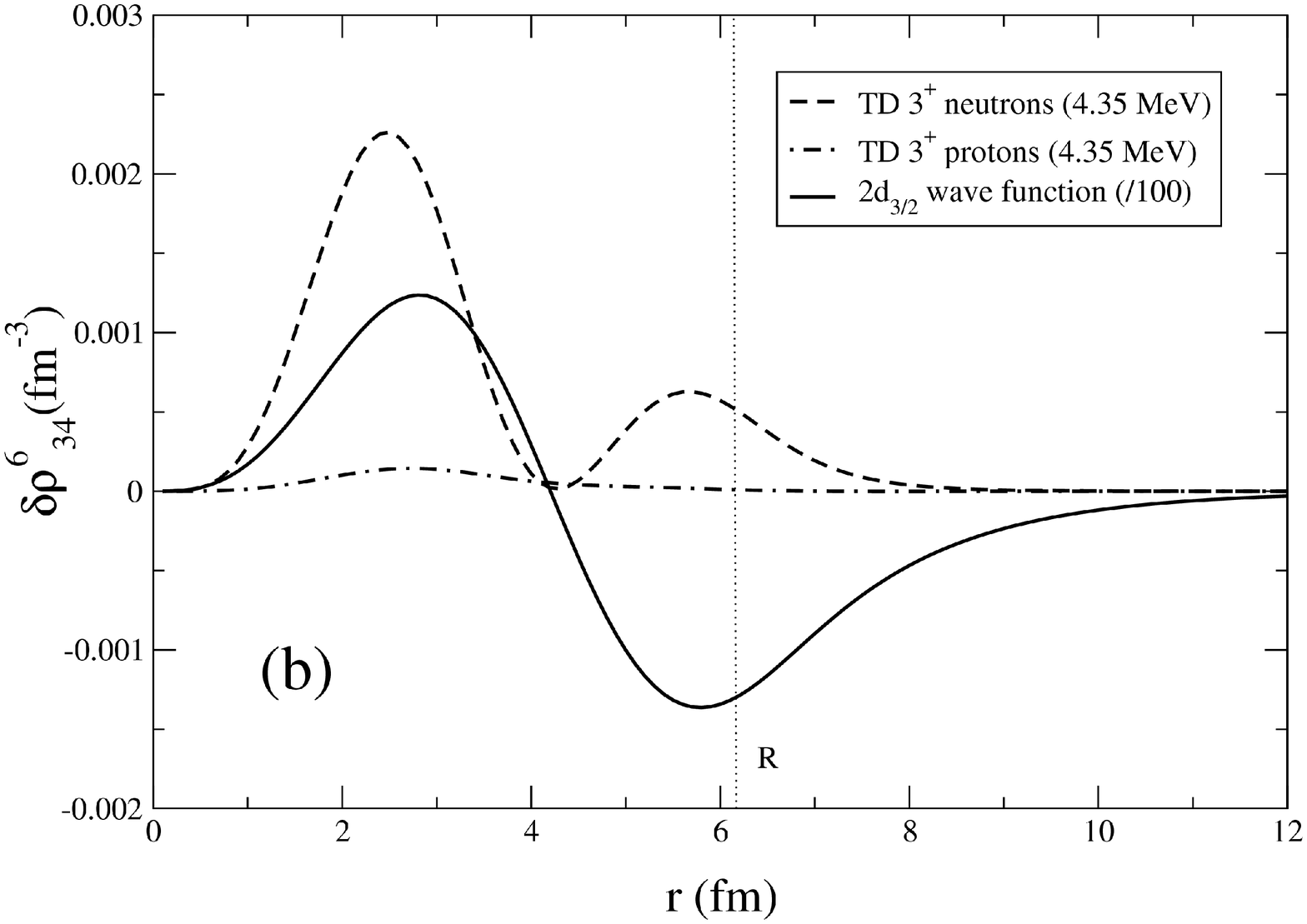,width=7.5cm}}
\caption{(a): the dashed and dot-dashed curves are respectively the neutron
and proton transition densities associated with the $2^+$ phonon
with energy 1.5 MeV while the solid curve is the wave function
of the 1h$_{11/2}$ state (in arbitrary units).
(b): same as (a) but for the $3^+$ phonon
with energy 4.35 MeV and the 2d$_{3/2}$ state.}
\label{fig:wave}
\end{figure}

To gain insight about the peculiar features of finite nuclei, 
as compared to the case of infinite systems, 
it is useful to study the radial dependence
of the particle-vibration coupling vertices shown in Fig. \ref{fig:fgraph}(b).
The induced pairing
matrix elements associated with natural parity modes 
have a clear surface character (cf. ref \cite{Hamamoto}). 
In particular 
this is the case for the most attractive 
pairing matrix element which is associated with
the 1h$^{2}_{11/2}$(0) ($\epsilon_{1h_{11/2}}=-8.07$ MeV, $\epsilon_F=-8.50$ MeV) configuration (cf. Fig. \ref{fig:diagonal}).
Because of its large centrifugal barrier, the
wave
function of this single-particle state is mainly concentrated at the nuclear surface. 
The main
contribution to the corresponding induced pairing matrix element arises from the exchange of a
$2^+$ phonon (of energy 1.5 MeV) between the two nucleons moving in time reversal
states in the h$_{11/2}$ orbital. The associated proton and
neutron transition densities depicted in Fig. \ref{fig:wave}(a) 
testify to the fact that this phonon has the character
of a surface vibration. Concerning the most 
repulsive matrix elements, we have found that the corresponding unnatural
parity phonons
are volume modes. In particular, one of the largest (positive) matrix element is associated
with the 2d$^2_{3/2}$(0) configuration ($\epsilon_{2d_{3/2}}$=-8.52 MeV). 
Because of the low angular
momentum, one finds that a consistent fraction of the corresponding 
wave function is concentrated in the interior of the nucleus. 
This state can thus couple efficiently with
phonons of volume character. In fact, the major contribution to
the corresponding matrix element is due to the exchange of the $3^+$
vibration (with energy at 4.35 MeV) which is a mode with a large volume component as testified by 
the corresponding proton and neutron transition densities shown in Fig. \ref{fig:wave}(b).
One can conclude that
states lying close to the Fermi energy with high $j$ and thus localized
at the surface mainly feel the (attractive)
coupling arising from the exchange of surface vibrations. 
The situation is expected to be quite different
in the case of infinite neutron matter. In fact, in going 
from the finite to the infinite system the collectivity
of the natural parity modes, mostly  surface modes, 
will be strongly reduced, while not much is expected to happen to the
volume modes.

\begin{figure}[b]
\centerline{\psfig{file=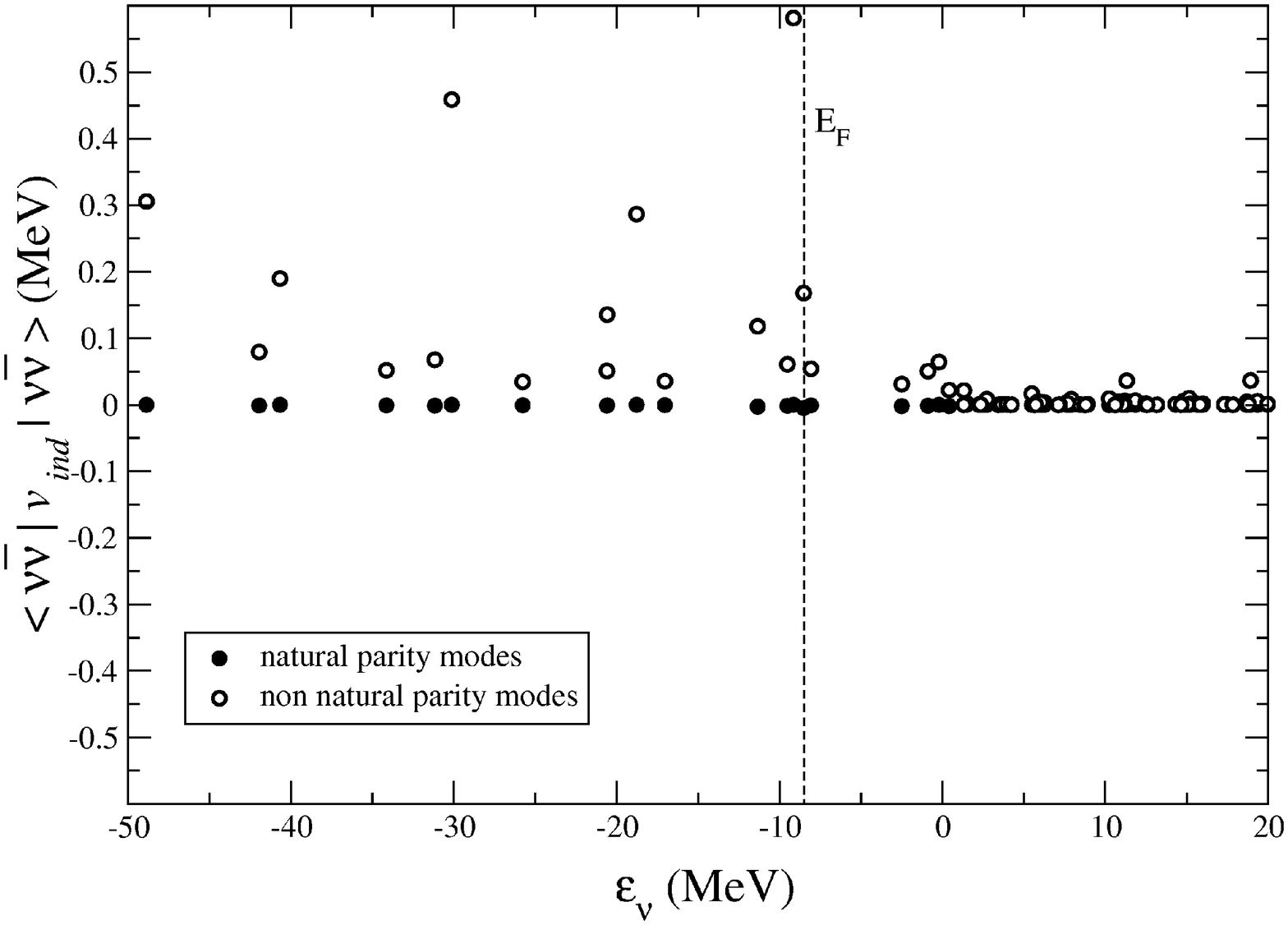,width=7.5cm}}
\caption{
The diagonal matrix elements produced by the exchange
of phonons with natural parity (filled circles) and those produced
by the exchange of phonons with non-natural parity (empty circles)
when the proton part of the phonon wave function is not included
in the calculation, are displayed as a function of the energy of
the single-particle states. 
} \label{fig:only}
\end{figure}

Furthermore, in going
from nuclear to neutron matter, many attractive contributions
vanish. In fact, if we turn off the neutron-proton particle-hole 
interaction contributing to
the  basic vertices displayed in Fig. \ref{fig:landp}, a strongly net repulsive
induced interaction is obtained (cf. Fig. \ref{fig:only}), a situation which much
resembles the neutron star case. 
This result can be understood by realizing that, quite generally,
 the  dominant  contribution to the  
spin independent (and therefore attractive) induced matrix elements arises 
precisely from 
the neutron-proton part of the particle-hole interaction, which is proportional to 
to   $(F_0 - F'_0) \delta \rho_p$ 
(cf. Eq. (4) and Fig. \ref{fig:landp}).  
The remaining part of the spin independent interaction depends on the 
function $F_0+F'_0$ (corresponding to the particle-phonon coupling mediated by 
$\delta \rho^i_{J^{\pi}n}$), which is rather weak, and 
for the SkM$^*$ interaction adopted here,  
even displays  a node at the nuclear surface. 
The induced interaction is then 
dominated by the (repulsive) spin-dependent matrix elements proportional to 
$G_0+G'_0$ (corresponding to the neutron-neutron particle-phonon coupling 
mediated by 
$\delta \rho^i_{J^{\pi}Ln}$, cf. Eq.(5)), which are large and without  nodes.

We have chosen to adopt a  
force like SkM*, which has been tested by different groups
over the years. We are well aware that our results could change adopting a different 
effective force, especially in connection with the fact that it is
difficult to fix the spin-dependent part of the
interaction, also because of the scarcity of experimental constraints.
In fact, different approaches have been adopted in the literature.
In particular, $1^+$ modes have been calculated in the framework of the extended theory of
finite Fermi systems, using constant values of the Landau parameters 
$G_0$ and $G'_0$  \cite{kamer,smirnov}.                            
Recently, a new specific parametrization of the Skyrme type has been 
introduced, SkO', which takes into account the effects of time-odd 
spin-isospin couplings, and has been adopted for the description 
of the spin-flip transitions like the Gamow-Teller resonance 
\cite{Engel,Bender}.
This force has not been yet
extensively checked in the non-charge-exchange channel, and 
the  large and negative value of $G_0$ associated with this force 
 can lead to too strongly collective or unstable solutions in the spin
isoscalar channel.
Even using different effective
forces, however, the main qualitative aspects of our results are likely to 
remain true, as they are based on quite general features 
- like the dominance of neutron-proton over neutron-neutron
interaction, and the stronger collectivity and surface localization of
the low-lying density modes with respect to the spin modes. 

We conclude that the exchange of low-lying surface vibrations (in which
neutrons and protons participate on equal footing) between pairs of 
nucleons moving in time reversal states close to the Fermi energy, leads to
a sizeable attractive pairing interaction which accounts for about 70\% of
the pairing gap. 
The inclusion of spin (volume) modes, reduces this contribution by 30\% in the
case of finite nuclei, bringing the 
induced pairing contribution to the pairing gap to a value of the order of
$\approx$50\%,
the other half coming from the bare nucleon-nucleon force.
The attractive character of the effective interaction in finite nuclei is 
found to be associated with the efficient coupling of the single-particle 
states lying close to the Fermi energy to collective surface vibrations,
as well as  with the contribution  of the proton-neutron part of the
particle-vibration coupling. Without these two elements, spin modes dominate and
the effective interaction becomes repulsive, as in neutron matter.



\bibliographystyle{apsrev}

\begin{thebibliography}{0}
\bibitem{note} To be noted that we here consider s-wave pairing, well aware 
of the fact that the exchange of collective modes can also be of relevance
in other channels (like p-wave pairing channels).
\bibitem{note1} Note that the induced matrix elements in this figure 
can be directly compared with the empirical constant G=25/A($\approx$0.2 MeV, A=120)
used to describe pairing correlations in nuclei within the framework of the BCS
theory and a (pairing) force of constant matrix elements.
\bibitem{addgap} The 
gap  obtained adding the total matrix elements of the induced 
interaction to those of the bare neutron-neutron interaction turns out to be 
substantially larger than that calculated with the bare interaction alone.
This is in contrast  with the neutron matter 
case and
in keeping with previous results for finite nuclei \cite{EPJ,Terasaki},
where, however, the spin-dependent particle-hole interaction 
had not been considered.
\end{thebibliography}

\end{document}